\title{On Locally Recoverable (LRC) Codes}
\author{Mario Blaum\\
IBM Almaden Research Center\\
San Jose, CA 95120
}
\date{}
\documentclass[12pt]{article}
\usepackage{latexsym}
\usepackage{multirow}
 \newtheorem{lemma}{Lemma}[section]

 \newtheorem{ex}{Example}[section]
 
 \newtheorem{cor}{Corollary}[section]
 
\newtheorem{COROLLARY}{\indent Corollary}

\newtheorem{EXAMPLE}{\indent Example}

\newtheorem{THEOREM}{\indent Theorem}

\newtheorem{REMARK}{\indent Remark}

\newcommand{\fullstop}{\hspace{-0.85em} {\bf .}}

\newcommand{\al}{\mbox{$\alpha$}}

\newcommand{\eq}{\mbox{$\, =\,$}}

\newcommand{\lan}{\mbox{$\langle$}}
\newcommand{\ran}{\mbox{$\rangle$}}

\newcommand{\qed}{\hfill$\Box$\\[1ex]}
\newcommand{\pf}{{\bf Proof: }}

\newcommand{\uzero}{\mbox{$\underline{0}$}}

\newcommand{\C}{\mbox{${\cal C}$}}

\newcommand{\br}{\\ }
\newcommand{\ce}{\begin{center}}
\newcommand{\cen}{\end{center}}

\newcommand{\ipb}{\begin{description}}
\newcommand{\ipn}{\end{description}}

\newcommand{\qb}{\begin{quote}}
\newcommand{\qn}{\end{quote}}

\newcommand{\tp}{\begin{titlepage}}
\newcommand{\tpn}{\end{titlepage}}
\newcommand{\zb}{\begin{figure}[hbtp]}
\newcommand{\zn}{\end{figure}}

\newcommand{\EQX}[1]{\begin{equation}\label{#1}}
\newcommand{\ENX}{\end{equation}}
\newcommand{\EQL}{\begin{eqnarray*}}
\newcommand{\ENL}{\end{eqnarray*}}
\newcommand{\EQLX}[1]{\begin{eqnarray}\label{#1}}
\newcommand{\ENLX}{\end{eqnarray}}

\newcommand{\open}{\begin{document}}
\newcommand{\close}{\end{document}}
\newcommand{\lfcr}[1]{\br\hspace*{#1em}}
\newenvironment{mat}[1]
{\left[\begin{array}{#1}}{\end{array}\right]}
\newcommand{\GAMMA}{\Gamma}
\newcommand{\DELTA}{\Delta}
\newcommand{\THETA}{\Theta}
\newcommand{\LAMBDA}{\Lambda}
\newcommand{\XI}{\Xi}
\newcommand{\PI}{\Pi}
\newcommand{\SIGMA}{\Sigma}
\newcommand{\UPSILON}{\Upsilon}
\newcommand{\PHI}{\Phi}
\newcommand{\PSI}{\Psi}
\newcommand{\OMEGA}{\Omega}
\newcommand{\bldgreek}[1]{\mbox{\boldmath $#1$}}
\newcommand{\bldbeta}{\bldgreek{\beta}}
\newcommand{\bldgamma}{\bldgreek{\gamma}}
\newcommand{\blddelta}{\bldgreek{\delta}}
\newcommand{\bldepsilon}{\bldgreek{\epsilon}}
\newcommand{\bldvarepsilon}{\bldgreek{\varepsilon}}
\newcommand{\bldzeta}{\bldgreek{\zeta}}
\newcommand{\bldeta}{\bldgreek{\eta}}
\newcommand{\bldtheta}{\bldgreek{\theta}}
\newcommand{\bldvartheta}{\bldgreek{\vartheta}}
\newcommand{\bldiota}{\bldgreek{\iota}}
\newcommand{\bldkappa}{\bldgreek{\kappa}}
\newcommand{\bldlambda}{\bldgreek{\lambda}}
\newcommand{\bldmu}{\bldgreek{\mu}}
\newcommand{\bldnu}{\bldgreek{\nu}}
\newcommand{\bldxi}{\bldgreek{\xi}}
\newcommand{\bldpi}{\bldgreek{\pi}}
\newcommand{\bldvarpi}{\bldgreek{\varpi}}
\newcommand{\bldrho}{\bldgreek{\rho}}
\newcommand{\bldvarrho}{\bldgreek{\varrho}}
\newcommand{\bldsigma}{\bldgreek{\sigma}}
\newcommand{\bldvarsigma}{\bldgreek{\varsigma}}
\newcommand{\bldtau}{\bldgreek{\tau}}
\newcommand{\bldupsilon}{\bldgreek{\upsilon}}
\newcommand{\bldphi}{\bldgreek{\phi}}
\newcommand{\bldvarphi}{\bldgreek{\varphi}}
\newcommand{\bldchi}{\bldgreek{\chi}}
\newcommand{\bldpsi}{\bldgreek{\psi}}
\newcommand{\bldomega}{\bldgreek{\omega}}
\begin{document}
\parindent=10pt
\maketitle
\begin{abstract}
We present simple constructions of optimal erasure-correcting LRC codes by exhibiting
their parity-check matrices. When the number of local parities
in a parity group plus the number of global parities is smaller than
the size of the parity group, the constructed codes are optimal with a
field of size at least the length of the code. We can reduce the size of
the field to at least the size of the parity groups when the number of
global parities equals the number 
of local parities in a parity group plus one.

\vspace{.3cm}

\noindent {\bf Keywords:} Erasure-correcting codes, Locally Recoverable codes,
Reed-Solomon codes, Generalized Concatenated codes, Integrated
Interleaved codes, Maximally Recoverable codes, MDS codes, Partial
MDS (PMDS)
codes, Sector-Disk (SD) codes, local and global parities, heavy parities.
\end{abstract}

\section{Introduction}
\label{Introduction}


Erasure-correcting codes combining local and global properties have
arisen considerable interest in recent
literature (see for instance~\cite{bhh}\cite{ghsy}\cite{hcl}\cite{hsx}\cite{pd}\cite{rk}\cite{sa}\cite{tb}\cite{twb}
and the references therein, a complete list of references on the
subject is beyond the scope of
this paper). The practical applications of these codes explain the reasons for this
interest. For example,
in storage applications involving multiple storage devices (like in
the cloud), most of the failures involve only one device. In that
case, it is convenient to recover the failed device ``locally,'' that
is, invoking a relatively small set of devices. This local recovery
process mitigates the decrease
in performance until the failed device is replaced. On the other hand, it
is desirable to have extra protection (involving global parities that
extend over all the devices) in case
more complex failure events occur. In this case, performance will be
impacted more severely (requiring even a halt in operations), but it
is expected that this is a rare event and when it occurs, data loss
is avoided. Examples
of these types of applications are Microsoft Windows Azure
Storage~\cite{hsx} and Xorbas
built on top of Facebook's Hadoop system running HDFS-RAID~\cite{sa}.

A different type of application consists of extending RAID
architectures in an array of storage devices to cover individual
sector or page failures in each storage device in the array.
In effect, certain storage devices like flash memories decay in time and with
use (number of writes). Every sector or page is protected by its own
Error-Correcting Code (ECC), but an accumulation of errors may exceed
the ECC capability of the sector. In that case, there is a ``silent
failure'' that will not be discovered until the failed sector or page
are accessed (scrubbing techniques are inconvenient for flash
memory, since they may impact future performance). For example, in
case an array of devices is protected 
using RAID 5, in the event a device fails, the sectors or pages in
the failed device are recovered by XORing the corresponding sectors
or pages
in the remaining devices. However, if one of the sectors or pages in
one of those remaining devices has failed (without a total device
failure), data loss will occur. Adding an extra parity device
(RAID~6) is expensive and moreover, may not correct some
combinations of failures (like two sector or page failures in a row
in addition to a device failure). This problem was studied 
in~\cite{bhh}\cite{bpsy}\cite{pb}.

Next we give some notation and define mathematically the problem.
Consider an erasure-correcting code $\C$ consisting of $m\times n$ arrays over a field
$GF(q)$. Each row in an array is
protected by $\ell$ local
parities as part of an $[n,n-\ell,\ell +1]$ MDS code, i.e., up to
$\ell$ erasures per row can be corrected. In 
addition, $g$ global parities are added. Then we say
that $\C$ is an $(m,n;\ell,g)$ Locally Recoverable (LRC) code.
Code $\C$ has length $mn$ and dimension
$k\eq m(n-\ell)-g$. As vectors, we consider the coordinates of the
arrays in $\C$ row-wise.

\begin{figure}
\centerline{
\begin{tabular}{|c|c|c|c|c|}
\hline
D&D&D&L&L\\\hline
D&D&D&L&L\\\hline
D&D&D&L&L\\\hline
D&G&G&L&L\\\hline
G&G&G&L&L\\\hline
G&G&G&L&L\\\hline
\end{tabular}
}
\caption{\label{fig0} A $6\times 5$ array showing the data symbols,
$\ell\eq 2$ local parities per row and $g\eq 8$ global parities.}
\end{figure}

The situation is illustrated in Figure~\ref{fig0} for a $6\times 5$
array with $\ell\eq 2$ local parities in each row and $g\eq 8$ global
parities. So, this is a $[30,10]$ code over a field $GF(q)$.

When an $(m,n;\ell,g)$ LRC code is desired, the key question is how
to construct the $g$ 
global parities in order to optimize the code. There are several
criteria for optimization and they depend on the application
considered. A weak optimization involves maximizing the
minimum distance of the code. To that end, bounds on the minimum distance are
needed.

Given two integers $s$ and $t\geq 1$, denote by $\lan s\ran_t$ the
(unique) integer $u$, $0\leq u\leq t-1$, such
that $u\equiv s\;(\bmod\;t)$. For example, $\lan 5\ran_3\eq 2$.
Also, given $x$, denote by $\lfloor x\rfloor$ the floor of $x$.
A Singleton type of bound on the minimum distance $d$ of an
$(m,n;\ell,g)$ LRC code is given
by~\cite{ghsy}

\begin{eqnarray}
\label{dist}
d&\leq & \ell +n\left\lfloor{g\over n-\ell}\right\rfloor +\lan g\ran_{n-\ell}+1.
\end{eqnarray}

For example, in Figure~\ref{fig0}, $\lfloor{g\over n-\ell}\rfloor\eq
2$ and $\lan g\ran_{n-\ell}\eq 2$, so, $d\leq 15$. This is easy to
see since, if any three rows are erased, 15 symbols are lost, but we
only have 14 parities to recover them, the 8 global parities and 6
local ones.
The general argument to prove bound~(\ref{dist}) proceeds similarly. We say that an LRC code is
optimal when bound~(\ref{dist}) is met with equality.

Let us point out that there are optimization criteria stronger
than the minimum distance. The
strongest one involves the so 
called Partial MDS (PMDS) codes~\cite{bhh}\cite{ghjy}
(in~\cite{ghjy}, PMDS codes are called Maximally Recoverable codes).
We say that an $(m,n;\ell,g)$ LRC code is PMDS if, whenever each row
of the code is punctured~\cite{ms} in exactly $\ell$ locations, 
the punctured code is MDS with respect to the $g$ parities (i.e., it
can correct any $g$ erasures after puncturing). A weaker
requirement (but still stronger than simply meeting bound~(\ref{dist})) is
provided by Sector-Disk (SD) codes~\cite{pb}:
these codes can correct any $g$ erasures after $\ell$ columns in an
array are erased (which corresponds to puncturing each row in $\ell$
locations, but these locations are fixed). Not surprisingly, PMDS and SD codes are  
harder to construct than optimal LRC codes. The known constructions
require a large field~\cite{bhh}\cite{ghjy}, although in some cases
efficient implementations (involving only XOR operations and
rotations of vectors) were found when the field is defined by the
irreducible polynomial $1+x+x^2+\cdots +x^{p-1}$, $p$ a prime and
$mn\,<\,p$~\cite{bhh}. Also, some good constructions 
of PMDS and SD codes (involving fields of size at least $2mn$ and $mn$
respectively) were found for $g\eq 2$~\cite{bpsy}, but the
general problem remains open.

When constructing optimal LRC codes, it is important to minimize the
size of the field.
Bound~(\ref{dist}), being a Singleton type of bound, does
not take into account this size. Most constructions of
optimal LRC codes in literature involved fields of large size, but
in~\cite{tb}, optimal LRC codes were obtained for fields of size
$q\geq mn$, that is, the length of the code, like in the case of
Reed-Solomon (RS) codes~\cite{ms}. In fact, it was shown in~\cite{tb}
that regular RS codes 
are a special case of the construction (when $m\eq 1$). For a bound on
LRC codes taking into account the size of the field, see~\cite{cm}.

In this paper, we examine a special case of $(m,n;\ell ,g)$ LRC codes that is important in
applications~\cite{hsx}\cite{sa}: the case $\ell +g\,<\,n$, hence, $\lfloor{g\over
n-\ell}\rfloor\eq 0$ and $\lan g\ran_{n-\ell}\eq g$, so bound~(\ref{dist})
becomes

\begin{eqnarray}
\label{dist1}
d&\leq & \ell +g+1.
\end{eqnarray}

\begin{figure}
\centerline{
\begin{tabular}{|c|c|c|c|c|}
\hline
D&D&D&L&L\\\hline
D&D&D&L&L\\\hline
D&D&D&L&L\\\hline
D&D&D&L&L\\\hline
D&D&D&L&L\\\hline
D&G&G&L&L\\\hline
\end{tabular}
}
\caption{\label{fig1} A $6\times 5$ array with $\ell\eq 2$ and $g\eq 2$.}
\end{figure}

For example, in Figure~\ref{fig1}, we have such a situation for a
$6\times 5$ array with $\ell\eq 2$ and $g\eq 2$, thus
bound~(\ref{dist1}) gives $d\leq 5$. The assumption $\ell +g\,<\,n$
allows us to obtain some very simple constructions.

The paper is organized as follows:
in Section~\ref{consgc}, in addition to the condition $\ell
+g\,<\,n$, we consider $(m,n;\ell ,g)$ LRC codes under the assumption
$g\leq \ell +1$. Then we 
use the techniques of Generalized Concatenated codes~\cite{bh1} to
obtain optimal $(m,n;\ell ,g)$ LRC codes over fields of size $q\geq
n$ (let us mention that there exist similar constructions in
literature for other applications, like magnetic recording~\cite{ah}\cite{hl}\cite{hapkt}\cite{pa}). 
In Section~\ref{cons} we present a simple construction over a field
of size $q\geq mn$ (as in~\cite{tb}) that had been given
in~\cite{bhh} as a candidate for PMDS codes, and we show that the
construction gives optimal $(m,n;\ell ,g)$ LRC codes when $\ell
+g\,<\,n$.  Finally, in Section~\ref{conc} we draw some conclusions justifying the 
assumption $\ell +g\,<\,n$.

\section{Construction of Optimal LRC Codes Using Generalized
Concatenation Techniques}
\label{consgc}
Denote by $RS(n,r;i,j)$, where $0\leq i$, $0\leq j$ and $1\leq r\,<\,n$, the following
parity-check matrix, corresponding
to a Reed-Solomon (RS) code of length $n$ and $r$ parities:

\begin{eqnarray}
\label{Hstn}
RS(n,r;i,j)&=&\left(
\begin{array}{ccccc}
\al^{ij}&\al^{i(j+1)}&\al^{i(j+2)}&\ldots &\al^{i(j+n-1)}\\
\al^{(i+1)j}&\al^{(i+1)(j+1)}&\al^{(i+1)(j+2)}&\ldots &\al^{(i+1)(j+n-1)}\\
\al^{(i+2)j}&\al^{(i+2)(j+1)}&\al^{(i+2)(j+2)}&\ldots &\al^{(i+2)(j+n-1)}\\
\vdots &\vdots &\vdots &\ddots &\vdots \\
\al^{(i+r-1)j}&\al^{(i+r-1)(j+1)}&\al^{(i+r-1)(j+2)}&\ldots &\al^{(i+r-1)(j+n-1)}\\
\end{array}
\right).
\end{eqnarray}

Also, denote by $I_m$ the $m\times m$ identity matrix and by $A
\otimes B$ the Kronecker product between matrices
$A$ and $B$~\cite{ms}.

Consider the $(m,n;\ell,g)$ LRC code over a field $GF(q)$
given by the
$(\ell m+g)\times mn$ matrix~\cite{bh1}

\begin{eqnarray}
\label{HGC}
H&=&\left(
\begin{array}{c}
I_m\otimes RS(n,\ell;0,0)\\
\hline
(\overbrace{1\;1\;\ldots\;1}^m)\otimes RS(n,g;\ell,0)\\
\end{array}
\right).
\end{eqnarray}

Then we have the following lemma:

\begin{lemma}
\label{l1}
{\em
Let $g\leq \ell+1$, $\ell +g\,<\,n$ and consider a field $GF(q)$ such
that $q\,>\, n$. Then the $(m,n;\ell,g)$ LRC code over $GF(q)$ whose
parity-check matrix $H$ is given by~(\ref{HGC}) is an optimal LRC code.
}
\end{lemma}

\pf Assume that $g\eq \ell+1$ (the cases $g\leq\ell$
are similar). We show that the code can correct any $g+\ell\eq 2\ell +1$ erasures.
In effect, assuming that $2\ell +1$ erasures have occurred, first we
correct all the rows having up to $\ell$ erasures using the $\ell$ local
parities in those rows. If any erasures are left after this process,
they are all in one row with at most $2\ell +1$ erasures. But from~(\ref{HGC}), there are
$2\ell +1$ parities
corresponding to a RS code ($\ell$ local parities and $\ell +1$
global parities) to correct these at most $2\ell +1$ erasures, so all
the erasures can be corrected since they involve inverting a $(2\ell
+1)\times (2\ell +1)$
Vandermonde matrix over the field $GF(q)$:
since $2\ell +1\eq\ell +g\,<\,n \,<\,q$, the elements in the
Vandermonde matrix are distinct and the matrix is invertible.
Thus, the minimum distance satisfies
$d\,\geq\, \ell +g+1$. But, since $\ell +g\,<\,n$, bound~(\ref{dist1})
gives $d\eq\ell +g+1$ and the code is optimal.

\qed

\begin{ex}
\label{ex1}
{\em
Take $m\eq 3$, $n\eq 6$, $\ell\eq 2$ and $g\eq 3$, then the
parity-check matrix over $GF(8)$ given by~(\ref{HGC}) is (notice, $\al^7\eq 1$)

\begin{eqnarray*}
H&=&\left(
\begin{array}{cccccc|cccccc|cccccc}
1&1&1&1&1&1&0&0&0&0&0&0&0&0&0&0&0&0\\
1&\al&\al^2&\al^3&\al^4&\al^5&0&0&0&0&0&0&0&0&0&0&0&0\\
0&0&0&0&0&0&1&1&1&1&1&1&0&0&0&0&0&0\\
0&0&0&0&0&0&1&\al&\al^2&\al^3&\al^4&\al^5&0&0&0&0&0&0\\
0&0&0&0&0&0&0&0&0&0&0&0&1&1&1&1&1&1\\
0&0&0&0&0&0&0&0&0&0&0&0&1&\al&\al^2&\al^3&\al^4&\al^5\\
\hline
1&\al^2&\al^4&\al^6&\al&\al^3&1&\al^2&\al^4&\al^6&\al&\al^3&1&\al^2&\al^4&\al^6&\al&\al^3\\
1&\al^3&\al^6&\al^2&\al^5&\al &1&\al^3&\al^6&\al^2&\al^5&\al
&1&\al^3&\al^6&\al^2&\al^5&\al \\
1&\al^4&\al &\al^5&\al^2&\al^6 &1&\al^4&\al &\al^5&\al^2&\al^6 &1&\al^4&\al &\al^5&\al^2&\al^6 \\
\end{array}
\right).
\end{eqnarray*}

It is easy to see directly that the parity-check matrix above allows for the
correction of any 5 erasures, so the minimum distance is 6 and the
code is optimal.

\qed
}
\end{ex}

We can obtain an optimal $(m,n;\ell,g)$ LRC code also for $q\,\eq\,n$ by
using extended RS codes, which requires a slight modification of
parity-check matrix~(\ref{HGC}). In effect, let

\begin{eqnarray}
\label{EHnrj}
ERS(n,r;j)&=&\left(
\begin{array}{cccccc}
1&1&1&\ldots &1&1\\
\al^{j}&\al^{j+1}&\al^{j+2}&\ldots &\al^{j+n-2}&0\\
\al^{2j}&\al^{2(j+1)}&\al^{2(j+2)}&\ldots &\al^{2(j+n-2)}&0\\
\al^{3j}&\al^{3(j+1)}&\al^{3(j+2)}&\ldots &\al^{3(j+n-2)}&0\\
\vdots &\vdots &\vdots &\ddots &\vdots&\vdots \\
\al^{(r-1)j}&\al^{(r-1)(j+1)}&\al^{(r-1)(j+2)}&\ldots &\al^{(r-1)(j+n-2)}&0\\
\end{array}
\right)
\end{eqnarray}
and consider the $(m,n;\ell,g)$ LRC code over a field $GF(q)$
given by the
$(\ell m+g)\times mn$ matrix

\begin{eqnarray}
\label{EHGC}
EH&=&\left(
\begin{array}{c}
I_m\otimes ERS(n,\ell;0)\\
\hline
(\overbrace{1\;1\;\ldots\;1}^m)\;\otimes\;
\left(RS(n-1,g;\ell,0)\;\;|\;\;\uzero_{g\times 1}\right)\\
\end{array}
\right).
\end{eqnarray}
where $\uzero_{i\times j}$ denotes an $i\times j$ zero matrix.

\begin{lemma}
\label{l11}
{\em
Let $g\leq \ell+1$, $\ell +g\,<\,n$ and take a field $GF(q)$ such
that $q\eq n$. Then the $(m,n;\ell,g)$ LRC code over $GF(q)$ whose
parity-check matrix $EH$ is given by~(\ref{EHGC}) is an optimal
$(m,n;\ell,g)$ LRC code.
}
\end{lemma}

\pf Completely analogous to the proof of Lemma~\ref{l1}, since an
extended RS code is also MDS.

\qed

\begin{ex}
\label{ex11}
{\em
Consider the situation described in~\cite{sa}. There, the authors
present a $(2,8;1,4)$ LRC code over $GF(16)$ with minimum distance 5.
Since $\ell +g\eq 5$, this code is not optimal. If we construct the
parity-check matrix $EH$ as given by~(\ref{EHGC}) for $m\eq 2$, $n\eq
8$, $\ell\eq 2$ and $g\eq 2$ over $GF(8)$, we obtain

\begin{eqnarray*}
EH&=&\left(
\begin{array}{cccccccc|cccccccc}
1&1&1&1&1&1&1&1&0&0&0&0&0&0&0&0\\
1&\al &\al^2&\al^3&\al^4&\al^5&\al^6&0&0&0&0&0&0&0&0&0\\
0&0&0&0&0&0&0&0&1&1&1&1&1&1&1&1\\
0&0&0&0&0&0&0&0&1&\al &\al^2&\al^3&\al^4&\al^5&\al^6&0\\
\hline
1&\al^2 &\al^4&\al^6&\al  &\al^3&\al^5&0&1&\al^2 &\al^4&\al^6&\al  &\al^3&\al^5&0\\
1&\al^3 &\al^6&\al^2&\al^5 &\al&\al^4&0&1&\al^3 &\al^6&\al^2&\al^{5}&\al&\al^4&0\\
\end{array}
\right).
\end{eqnarray*}

By Lemma~\ref{l11}, this code has minimum distance 5 and is an optimal
LRC code. Compared to the code in~\cite{sa}, the minimum distance is
the same, but this code can correct up to two erasures per row and
it requires a smaller field, i.e., $GF(8)$ as opposed to $GF(16)$.
However, the locality in~\cite{sa}, that is, the number of symbols necessary
to reconstruct a lost (data) symbol, is 5, while in this example it
is 6. Moreover, it was proven in~\cite{sa} that
if the locality is 5, the minimum distance is at most 5, so the code
in~\cite{sa} satisfies a different type of optimality. The
decision on which code to use and which type of optimality is
preferable depends on the application, there are trade-offs that need
to be considered.

\qed
}
\end{ex}

\section{Construction of Optimal LRC Codes when $\ell +g\,<\,n$}
\label{cons}

Consider Construction~3.2 in~\cite{bhh}. For the sake of completeness, we
present it next.
Let the $(m,n;\ell,g)$ LRC code over $GF(q)$, $q\,>\, mn$, be given by the parity-check matrix

\begin{eqnarray}
\label{HLRC}
H(m,n;\ell,g)&=&\left(
\begin{array}
{c|c|c|c}
RS(n,\ell;0,0)&\uzero_{\ell\times n}&
\ldots &\uzero_{\ell\times n}\\
\uzero_{\ell\times n}&RS(n,\ell;0,r)&
\ldots &\uzero_{\ell\times n}\\
\vdots & \vdots & 
\ddots &\vdots\\
\uzero_{\ell\times n}&\uzero_{\ell\times n}&
\ldots &RS(n,\ell;0,(m-1)\ell)\\
\hline
\multicolumn{4}{c}{RS(mn,g;\ell,0)}\\
\end{array}
\right),
\end{eqnarray}

\begin{ex}
\label{ex2}
{\em
Take $m\eq 3$, $n\eq 5$, $\ell\eq 1$, $g\eq 3$ and the field $GF(16)$.
According to~(\ref{HLRC}), we obtain

\begin{eqnarray*}
H(3,5;1,3)&=&\left(
\begin{array}{ccccc|ccccc|ccccc}
1&1&1&1&1&0&0&0&0&0&0&0&0&0&0\\
0&0&0&0&0&1&1&1&1&1&0&0&0&0&0\\
0&0&0&0&0&0&0&0&0&0&1&1&1&1&1\\
\hline
1&\al&\al^2&\al^3&\al^4&\al^5&\al^6&\al^7&\al^8&\al^9&\al^{10}&\al^{11}&\al^{12}&\al^{13}&\al^{14}\\
1&\al^2&\al^4&\al^6&\al^8&\al^{10}&\al^{12}&\al^{14}&\al&\al^3&\al^{5}&\al^{7}&\al^{9}&\al^{11}&\al^{13}\\
1&\al^3&\al^6&\al^9&\al^{12}&1&\al^3&\al^6&\al^9&\al^{12}&1&\al^3&\al^6&\al^9&\al^{12}\\
\end{array}
\right).
\end{eqnarray*}

Similarly, if $m\eq 3$, $n\eq 5$, $\ell\eq 2$ and $g\eq 2$,
(\ref{HLRC}) gives

\begin{eqnarray*}
H(3,5;2,2)&=&\left(
\begin{array}{ccccc|ccccc|ccccc}
1&1&1&1&1&0&0&0&0&0&0&0&0&0&0\\
1&\al&\al^2&\al^3&\al^4&0&0&0&0&0&0&0&0&0&0\\
0&0&0&0&0&1&1&1&1&1&0&0&0&0&0\\
0&0&0&0&0&\al^5&\al^6&\al^7&\al^8&\al^9&0&0&0&0&0\\
0&0&0&0&0&0&0&0&0&0&1&1&1&1&1\\
0&0&0&0&0&0&0&0&0&0&\al^{10}&\al^{11}&\al^{12}&\al^{13}&\al^{14}\\
\hline
1&\al^2&\al^4&\al^6&\al^8&\al^{10}&\al^{12}&\al^{14}&\al&\al^3&\al^{5}&\al^{7}&\al^{9}&\al^{11}&\al^{13}\\
1&\al^3&\al^6&\al^9&\al^{12}&1&\al^3&\al^6&\al^9&\al^{12}&1&\al^3&\al^6&\al^9&\al^{12}\\
\end{array}
\right).
\end{eqnarray*}

\qed
}
\end{ex}

We have the following lemma:

\begin{lemma}
\label{l2}
{\em
Consider the $(m,n;\ell,g)$ LRC code over $GF(q)$ with $q\,>\, mn$ whose
parity-check matrix $H(m,n;\ell,g)$ is given by~(\ref{HLRC}). Then
the code
has minimum distance $d\eq\ell +g+1$.
}
\end{lemma}

\pf
Notice that if we take the linear combination of the rows of
$H(m,n;\ell,g)$ consisting of XORing rows
$0,\ell,2\ell,\ldots,(m-1)\ell$, then rows
$1,\ell+1,2\ell+1,\ldots,(m-1)\ell+1$, and so on, up to rows
$\ell-1,2\ell-1,3\ell-1,\ldots,m\ell-1$, we obtain the
$(\ell+g)\times mn$ matrix $RS(mn,\ell+g;0,0)$.

Consider any $(\ell m+g)\times (\ell +g)$ submatrix of
$H(m,n;\ell,g)$. We have to prove that this matrix has rank $\ell
+g$, which is at least the rank of any linear
combination of its rows. In particular, it is at least the rank of
the corresponding $(\ell +g)\times (\ell +g)$
submatrix of $RS(mn,\ell+g;0,0)$, which is $\ell +g$ since it is a
Vandermonde matrix and $mn\,<\,q$.

\qed

\begin{cor}
\label{co1}
{\em
Consider an $(m,n;\ell,g)$ LRC code with the conditions of
Lemma~\ref{l2} and assume that $\ell +g\,<\,n$. Then the code is an
optimal LRC code.
}
\end{cor}

\pf Simply notice that, by Lemma~\ref{l2}, inequality~(\ref{dist1})
is met with equality.

\qed

\begin{ex}
\label{ex3}
{\em Consider the $(3,5;2,2)$ LRC code whose
parity-check matrix is $H(3,5;2,2)$ as given in Example~\ref{ex2}.
According to Lemma~\ref{l2}, its
minimum distance is 5. To view this, XORing rows 0, 2 and 4, and then
rows 1, 3 and 5 of $H(3,5;2,2)$, the $4\times 15$ matrix

\begin{eqnarray*}
RS(15,4;0,0)&=&\left(
\begin{array}{ccccc|ccccc|ccccc}
1&1&1&1&1&1&1&1&1&1&1&1&1&1&1\\
1&\al&\al^2&\al^3&\al^4&\al^5&\al^6&\al^7&\al^8&\al^9&\al^{10}&\al^{11}&\al^{12}&\al^{13}&\al^{14}\\
\hline
1&\al^2&\al^4&\al^6&\al^8&\al^{10}&\al^{12}&\al^{14}&\al&\al^3&\al^{5}&\al^{7}&\al^{9}&\al^{11}&\al^{13}\\
1&\al^3&\al^6&\al^9&\al^{12}&1&\al^3&\al^6&\al^9&\al^{12}&1&\al^3&\al^6&\al^9&\al^{12}\\
\end{array}
\right)
\end{eqnarray*}
is obtained. Any $8\times 4$ submatrix of $H(3,5;2,2)$ has rank 4,
since the corresponding $4\times 4$ submatrix of $RS(15,4;0,0)$ also
has rank 4.

\qed
}
\end{ex}

In Lemma~\ref{l2}, we have shown how to construct $(m,n;\ell,g)$ LRC
codes over $GF(q)$ with $q\,>\, mn$ and minimum distance $d\eq \ell
+g+1$. We can also construct a code with these parameters and
$q\eq mn$ by considering extended RS codes as done in
Section~\ref{consgc}.
Using~(\ref{EHnrj}) consider the $(m,n;\ell,g)$ LRC code over
$GF(q)$, $q\eq mn$, given by the parity-check matrix

\begin{eqnarray}
\label{EHLRC}
EH(m,n;\ell,g)&=&\left(
\begin{array}
{c|c|c|c}
RS(n,\ell;0,0)&\uzero_{\ell\times n}&
\ldots &\uzero_{\ell\times n}\\
\uzero_{\ell\times n}&RS(n,\ell;0,r)&
\ldots &\uzero_{\ell\times n}\\
\vdots & \vdots & 
\ddots &\vdots\\
\uzero_{\ell\times n}&\uzero_{\ell\times n}&
\ldots &ERS(n,\ell;(m-1)\ell)\\
\hline
\multicolumn{4}{c}{RS(mn-1,g;\ell,0)\quad|\quad\uzero_{g\times 1}}\\
\end{array}
\right).
\end{eqnarray}

\begin{lemma}
\label{l3}
{\em
The $(m,n;\ell,g)$ LRC code over $GF(q)$ with $q\eq mn$ whose
parity-check matrix $EH(m,n;\ell,g)$ is given by~(\ref{EHLRC})
has minimum distance $d\eq\ell +g+1$.
}
\end{lemma}

\pf
Completely analogous to the proof of Lemma~\ref{l2}, except that
after making the XORs described there we obtain the $(\ell+g)\times
mn$ matrix $ERS(mn,\ell+g;0)$, corresponding to an extended RS code
and which is MDS.

\qed

\begin{ex}
\label{ex4}
{\em
Consider again the situation of a $(2,8;1,4)$ LRC code over $GF(16)$
with minimum distance 5 described in~\cite{sa}. If we construct
$EH(2,8;1,4)$ also over $GF(16)$ as given by~(\ref{EHLRC}), we obtain

\begin{eqnarray*}
EH(2,8;1,4)&=&\left(
\begin{array}{cccccccc|cccccccc}
1&1&1&1&1&1&1&1&0&0&0&0&0&0&0&0\\
0&0&0&0&0&0&0&0&1&1&1&1&1&1&1&1\\
\hline
1&\al
&\al^2&\al^3&\al^4&\al^5&\al^6&\al^7&\al^8&\al^9&\al^{10}&\al^{11}&\al^{12}&\al^{13}&\al^{14}&0\\
1&\al^2 &\al^4&\al^6&\al^8&\al^{10}&\al^{12}&\al^{14}&\al&\al^{3}&\al^{5}&\al^{7}&\al^{9}&\al^{11}&\al^{13}&0\\
1&\al^3
&\al^6&\al^9&\al^{12}&1&\al^{3}&\al^{6}&\al^9&\al^{12}&1&\al^3
&\al^6&\al^9&\al^{12}&0\\
1&\al^4 &\al^8&\al^{12}&\al &\al^{5}&\al^{9}&\al^{13}&\al^2&\al^{6}&\al^{10}&\al^{15}&\al^{3}&\al^{7}&\al^{11}&0\\
\end{array}
\right).
\end{eqnarray*}

By Lemma~\ref{l3}, this code has minimum distance 6 and hence is an optimal
LRC code. The locality, though, is 7, as opposed to 5 in~\cite{sa}.

\qed
}
\end{ex}

\section{Conclusions}
\label{conc}
We have studied the problem of $(m,n;\ell,g)$ LRC codes and presented some simple
constructions for the case $\ell +g\,<\,n$. The size of the field is
$q\geq mn$ and in some cases (when in addition, $g\leq\ell +1$), $q\geq n$.

The general case without the restriction $\ell +g\,<\,n$ is handled
in~\cite{tb} and the construction there is much more involved. For
example, consider $6\times 5$ arrays like in Figures~\ref{fig0}
and~\ref{fig1}. In Figure~\ref{fig1} with $\ell\eq g\eq 2$,
an optimal LRC code with minimum distance $d\eq 5$ is obtained
with the construction presented. However, if $g\eq 3$, the minimum
distance is $d\eq 6$, while the optimal, according to
bound~(\ref{dist}), is $d\eq 7$. The remarkable construction
in~\cite{tb} effectively provides such a code. So, the question is,
how relevant is the condition $\ell +g\,<\,n$
in applications?

A channel requiring a large number
of global parities $g$ may neglect the impact of local
recovery. As stated in the Introduction, it is desirable that the
local parities are the ones most 
heavily used. The reason they have been introduced is due to
a performance issue: they allow for fast recovery in case 
the number of erasures in a row does not exceed $\ell$. The global
parities are used in case the erasure-correcting power of the local
parities is exceeded, but for
the system to be efficient, these should be relatively rare cases. If the
global parities need to be frequently invoked, we may do better with
an MDS code (saving in parity), or, with the same amount of parity,
having a much stronger code. For example, if we take the $[30,10]$
code of Figure~\ref{fig0}, in the optimal case we can correct 14
erasures. But if we used an MDS code we could correct 20 erasures by
sacrificing the locality. Alternatively, we could use a $[30,16]$ MDS
code that can also correct up to 14 erasures, but with a much better rate.

\end{document}